\documentstyle[aps,epsfig]{revtex}

\font\mb=msbm10

\begin{document}

\newcommand{\mt}{\mu_{t}}
\newcommand{\md}{{\bf{M}}^{-\tau}}
\newcommand{\ds}{\Delta_{\rm i}^{\tau}S}
\newcommand{\sk}{s_{k}}
\newcommand{\vrr}{\vec{r}}
\newcommand{\cM}{{\cal{M}}}
\newcommand{\vL}{\vec{L}\left[{\Phi}^{-t}(\vl,X,0)\right]}
\newcommand{\cb}{\cal{B}}
\newcommand{\re}{\rho_{\rm eq}}
\newcommand{\intk}{{\frac{1}{|\cal{B}|}}\int_{\cal{B}}\,d\vec{k}\;F_{\vk}}
\newcommand{\intknu}{{\frac{\nu(\cM)}{|\cal{B}|}}\int_{\cal{B}}\,d\vec{k}\;
F_{\vk}}
\newcommand{\dlt}{\vec{d}(X,t)}
\newcommand{\A}{\{ A_i \}}
\newcommand{\ct}{{\cal{T}}}
\newcommand{\vk}{\vec{k}}
\newcommand{\vl}{\vec{l}}
\newcommand{\vp}{\vec{p}}
\newcommand{\vv}{\vec{v}}
\newcommand{\vg}{\vec{g}}
\newcommand{\be}{\begin{equation}}
\newcommand{\ee}{\end{equation}}
\newcommand{\bea}{\begin{eqnarray}}
\newcommand{\eea}{\end{eqnarray}}
\newcommand{\G}{X}

\newtheorem{definition}{Definition}
\newtheorem{lemma}{Lemma}[definition]

\draft


\title{Diffusive Lorentz gases and multibaker maps \\
are compatible with irreversible thermodynamics}

\author{P. Gaspard$^{1}$, G. Nicolis$^{1}$, and J. R. Dorfman$^{2}$}
\address{$^{1}$ Center for Nonlinear Phenomena and Complex Systems, 
Universit\'e Libre de Bruxelles, \\
Campus Plaine, Code Postal 231, 
B-1050 Brussels, Belgium \\
$^{2}$ Department of Physics and Institute for Physical Science and
Technology, \\
University of Maryland, 
College Park, MD 20742, USA}


\maketitle


\begin{abstract}
We show that simple diffusive systems, such as the Lorentz gas and
multibaker maps are perfectly compatible with the laws of irreversible
thermodynamics, despite the fact that the moving particles, or their
equivalents, in these models do not
interact with each other, and that the dynamics takes place in
low-dimensional phase spaces.
The interaction of moving particles with scatterers
provides the dynamical mechanism responsible for an approach to
equilibrium, under appropriate conditions. This analysis provides
a refutation of the criticisms expressed recently by Cohen and Rondoni
[Physica A {\bf 306} (2002) 117-128].
\end{abstract}

\pacs{{\it PACS:} 05.20.-y; 05.45.+b; 05.60.+w; 05.70.Ln\\ \\
{\it Keywords:} Nonequilibrium statistical mechanics; Irreversible
thermodynamics; Dynamical systems}

\maketitle


\section{Introduction}

For nearly one hundred years, the Lorentz gas has been investigated as a
model of diffusive transport of light
tracer particles among heavier particles \cite{Lo05,vBe82,De00,CC70,Ha74}.
On the collisional time scale of the fast tracer
particles, the heavy particles can be treated immobile, which
simplifies the dynamics.  Due to this
simplification, the Lorentz gas plays a privileged and important role in
the development of transport and kinetic
theories.  In the last decades, several versions
of the Lorentz gas model have been
mathematically studied in detail and the
existence of a well-defined diffusion coefficient has been proved
rigorously under
certain conditions \cite{BuSi80,Kn87}.
Moreover, the Einstein relation between the coefficients of electrical
conductivity
and diffusion has also been proved for a Gaussian thermostated Lorentz gas
in the presence of an external field
\cite{ChEyLeSi93a,ChEyLeSi93b}.

Ten years ago, one of us
introduced the multibaker map as a simple
model of deterministic
diffusion \cite{Ga92}.  This map is
constructed by simplifying the Birkhoff-Poincar\'e map of the hard-disk
Lorentz gas.  The multibaker map is a
spatial extension of the baker map
described by Arnold and
Avez \cite{ArAv68} and previously by
Hopf \cite{Hopf37} who also described a
spatially extended baker-type model identical, in many respects,
to the multibaker map.\footnote{ The baker map was introduced in the
thirties in the work by Seidel \cite{Seidel33} who
seems to have been inspired by Birkhoff.} The advantage of the multibaker
map for a study of deterministic diffusion is that, like the baker map,
it is exactly solvable and can be used to test specific hypotheses in the
context of kinetic and transport theories.
Several versions of the multibaker map have been proposed and investigated
as models of different transport
processes
\cite{Ga92,Ga98,TeVoBr96,TaGa95,TaGa99,TaGa00,Ra99,GiDo99,GiFeDo99,GiDo00,TeVo00,VoTeMa00,TeVoMa01,MaTeVo01,Vo02}.

Recently Cohen and Rondoni have written a series of papers \cite{CRNL,CoRo02,CRPD}
where, among other things, the physical relevance of the Lorentz gas and multibaker map
in irreversible thermodynamics is
challenged.  Their main criticism is that the Lorentz gas and multibaker map
``represent noninteracting particle
systems" so that they ``do not possess the crucial property of local
thermodynamic equilibrium".  Hence ``these
models are not suited for a derivation of irreversible thermodynamics"
(Ref. \cite{CoRo02} p. 117) where, in their view, ``the local physical
density gradient $\partial n/\partial x$ acts like a real force which does
work on the particles themselves".  In contrast, as claimed,
``in the Lorentz slab there is no dissipation, no work done on the
moving particles, hence no entropy
production" (Ref. \cite{CoRo02} p. 127).

Our purpose in this paper is to refute these criticisms.
We show that
the views of Cohen and Rondoni concerning
both the status of the density gradient as a real force
performing work on the particles and the physical unsoundness
of the Lorentz gas and multibaker
map from the point of view of irreversible thermodynamics, are
wrong. M\'aty\'as, T\'el, and Vollmer have recently responded to
criticisms of Cohen and Rondoni, directed at their work as well as
ours, and expressed views along lines
similar to those given here \cite{MaTeVo02}.

The plan of the paper is the following.
In Section \ref{thermo},
we give a brief review of the irreversible thermodynamics
of diffusion in binary mixtures and clarify the status of the
thermodynamic forces present.
In the following Sections
\ref{Lorentz} and \ref{multibaker}, we show that diffusion in the
Lorentz gas and in the
multibaker map is well described by, and compatible with
irreversible thermodynamics, and that a local equilibrium
indeed establishes itself in these systems.


\section{The irreversible thermodynamics of matter exchanges}
\label{thermo}

In this section, we provide a first response to the claim that
thermodynamic forces do
work on the particles {\it themselves} and
that this is a {\it sine qua non} condition for irreversible entropy
production. We briefly review the irreversible thermodynamics of
isothermal diffusion in multi-component mixtures and show how it applies
to Lorentz gases, considered as binary mixtures of light and
infinitely massive particles.  Moreover, we show that the entropy production
of diffusion does not vanish in binary mixtures of particles with
an extreme mass ratio.

\subsection{The Second Law in the presence of matter exchanges}

Classical irreversible thermodynamics is a local theory.
It rests on the sole assumption that the densities of entropy $s$,
internal energy $e$, and particle numbers $n_k$ are locally linked
through a relation having the same structure as the relation
linking the global analogs of these quantities in equilibrium:
\be
s(\vec{r},t) = s\left[ e(\vec{r},t), \{n_k(\vec{r},t)\}\right] \; .
\label{s}
\ee
The differentiable function $s$ has the same properties
as the entropy
\be
S=S(E,V,\{N_k\})
= V \; s\left(\frac{E}{V},\left\{\frac{N_k}{V}\right\}\right) \; ,
\label{S1}
\ee
of a system at the internal energy $E$ and
containing $N_k$ particles of substance
$k$ in a volume $V$.  Here, the identity of the
two differentiable forms allows one
to interpret the derivatives of $s$ with respect to $e$ and $n_k$
as being, respectively, the local reciprocal temperature
and minus the chemical potential of substance $k$ multiplied
by the reciprocal temperature.

Starting from Eq. (\ref{s}) one can derive a balance equation
of entropy by differentiating both sides with respect to time
and by using the balance equations of mass, momentum,
and energy of macroscopic physics.
One obtains then in the balance equation a source term --
the entropy production -- in the form
\be
\sigma_{\rm irr} = \sum_{\alpha} J^{(\alpha)} X^{(\alpha)} \geq 0 \; ,
\label{irr}
\ee
where $J^{(\alpha)}$ is the thermodynamic flux associated with
process $\alpha$ and $X^{(\alpha)}$ is the conjugate thermodynamic force.
In a given system all these processes need not coexist.  When they do
they are, typically, coupled as long as certain symmetry requirements
are satisfied.

The status of Eq. (\ref{s}) from the standpoint of microscopic theory
has been the subject of many investigations.  Within the context of
kinetic theory of gases, Prigogine \cite{Pr49} showed that a
sufficient condition of validity is that the velocity distribution
must remain close to the Maxwellian.  This guarantees, at the same time,
the validity of linear phenomenological relations linking the
fluxes to the forces, a property that is less general
than Eq. (\ref{irr}).
However, Eq. (\ref{s}) may still be correct under circumstances
where the local equilibrium distribution is not a near-Maxwellian
distribution in the energy \cite{NiWaVe69}.
The near-Maxwellization is sufficient for the validity of (\ref{s})
but does not constitute a necessary condition.
As a matter of fact, the main physics behind Eq. (\ref{s}) is that
after a short initial relaxation period the space-time variation
of the probability distribution descriptive of the
nonequilibrium system at hand can be expressed entirely in terms of
the variation of the macroscopic (hydrodynamic) fields.

If a system exchanges matter $d{\cal N}_k$ as well as heat $dQ$ with its
surroundings, the Second Law reads \cite{Haase69}
\be
\Delta S = S_{\rm II}-S_{\rm I} \geq \int_{\rm I}^{\rm II}
\left(\frac{dQ}{T} + \sum_k {\cal S}_k\; d{\cal N}_k \right) \; ,
\label{general}
\ee
where $\Delta S$ is the total change of entropy in the system, not
including the surroundings, ${\cal S}_k$ is the partial molar entropy of
substance $k$,
and ${\cal N}_k$ denotes the number of moles of this substance.  In a
closed system that
does not exchange matter with its surrounding (i.e., $d{\cal N}_k=0$) but
only heat, we
recover the well-known inequality
\be
\Delta S = S_{\rm II}-S_{\rm I} \geq \int_{\rm I}^{\rm II} \frac{dQ}{T} \; ,
\ee
which implies that if no heat is transferred to a closed
system, there may still be a transient irreversible increase of entropy.
On the other hand, in an open system,
matter exchange contributes to entropy
production beside processes of heat exchange.
In this regard, we mention
that the study of processes of exchange of
matter requires the consideration of chemiostats (particle reservoirs)
instead of thermostats (heat reservoirs).

\subsection{Irreversible thermodynamics of diffusive processes in mixtures}

Transport by diffusion is a process in which there is an exchange of
matter from one macroscopic part of a system to another.  This
irreversible process can take place in a system in which
no other irreversible process occurs.  We then have a process where
there is an exchange of matter contributing to entropy production
without further contribution, for instance,
due to temperature or macroscopic velocity gradients.  Processes of
exchange of matter are very common in
chemical thermodynamics \cite{BeRiRo80}.
In particular, the process of mixing of two distinct chemical
species is well known to contribute to entropy production.
In such processes, spatial
inhomogeneities of chemical concentrations exist
across the system before the thermodynamic equilibrium is reached.  The
diffusion of tracer particles in a fluid
of other particles is an example of such an irreversible process with
exchange of matter.

A fluid composed of tracer particles moving among other particles can be
considered as a binary mixture.  It is well known that an isothermal
$c$-component mixture at rest
(i.e., with a zero macroscopic velocity field) can
only sustain $(c-1)$ independent diffusive processes \cite{Haase69,dGrMa62}.
In a binary mixture, we should thus expect a single diffusive process.
It is a process of mutual diffusion between the tracer particles
and the other particles.  The position of the tracer particles
is defined with respect to the center of mass of the total system.

The thermodynamics of a $c$-component mixture is based on the Gibbs
relation
\be
dE=T \; dS-P \;  dV + \sum_{k=1}^c \; \mu_k \; dN_k \; ,
\ee
where $T$ is the temperature, $P$ the pressure,
and $\mu_k$ the chemical potential of particles $k$.
If we define the densities of energy, entropy, and particles $k$
respectively as [cf. Eq. (\ref{S1})]
\be
e = \frac{E}{V},  \qquad  s=\frac{S}{V} \; , \qquad \mbox{and}\qquad
n_k=\frac{N_k}{V} \; ,
\label{densities}
\ee
we obtain the local Gibbs relation \cite{KoPr98}
\be
de = T \; ds + \sum_{k=1}^c \mu_k \; dn_k \; .
\label{local.Gibbs}
\ee
The extensivity  of the energy, entropy, and particle numbers
together with the local Gibbs relation (\ref{local.Gibbs}) imply
the Gibbs-Duhem relation
\be
s \; dT - dP + \sum_{k=1}^c n_k \; d\mu_k = 0 \; ,
\ee
which shows that, for an isothermal and isobaric system ($dT=dP=0$),
the variations of the chemical potentials are interconnected by
\be
\sum_{k=1}^c n_k \; \vec{\nabla}\mu_k = 0 \; .
\label{iso.Gibbs.Duhem.grad}
\ee

On the other hand, in absence of chemical reactions,
local conservation of particles holds, yielding
\be
\partial_t \; n_k + \vec{\nabla}\cdot\left( n_k \,\vec{u} + \vec{J}_k \right)
= 0
\; ,
\label{loc.cons.nber}
\ee
where $\vec{J}_k$ is the current density of particles $k$ with respect
to the local barycentric velocity $\vec{u}$ of the fluid,
which is the velocity of the center
of mass of a fluid element.
The local conservation of the total mass implies that
the currents are related by
\be
\sum_{k=1}^c m_k \; \vec{J_k} = 0 \; ,
\label{mass.constraint}
\ee
where $m_k$ denotes the mass of one particle of component $k$.

As a consequence of the constraints
(\ref{iso.Gibbs.Duhem.grad}) and (\ref{mass.constraint})
only $(c-1)$ gradients and currents are
independent because
\begin{eqnarray}
\vec{\nabla}\mu_c &=& - \sum_{k=1}^{c-1} \frac{n_k}{n_c}\; \vec{\nabla}\mu_k
\; , \\
\vec{J}_c &=& - \sum_{k=1}^{c-1} \frac{m_k}{m_c}\; \vec{J}_k \; .
\end{eqnarray}

In an isothermal $c$-component mixture at rest ($\vec{u}=0$), the entropy
production is therefore given by \cite{dGrMa62}
\be
\sigma_{\rm irr} = \sum_{k=1}^c \vec{J}_k \cdot \left( -
\vec{\nabla}\frac{\mu_k}{T}\right)
 = \sum_{k=1}^{c-1} \vec{J}_k \cdot \vec{X}_k \; ,
\ee
with the thermodynamic forces
\be
\vec{X}_k \equiv - \vec{\nabla}\frac{\mu_k}{T} - \sum_{l=1}^{c-1}
\frac{m_kn_l}{m_cn_c} \; \vec{\nabla}\frac{\mu_l}{T} \; .
\ee
In a binary mixture ($c=2$), the entropy production contains only the
contribution from the single diffusive process the system can
sustain\footnote{Equation (\ref{entr.prod.mutual.diff}) is given by de Groot
and Mazur \cite{dGrMa62} as Eq. (94) of Chapter XI in the form
$\sigma_{\rm irr} = - \frac{1}{a_2}  \vec{J}_1 \cdot
\vec{\nabla}\frac{\mu_1}{T}$ where $a_2$ is the mass fraction of substance
$2$ which is indeed given by
$a_2=m_2n_2/(m_1n_1+m_2n_2)$.  In the Lorentz-gas limit $(m_1/m_2)\to 0$,
the mass fraction
of substance 2 equals $a_2=1$. We notice that we are here using quantities
per unit volume
while de Groot and Mazur use quantities per unit mass in Ref.
\cite{dGrMa62}.  Both derivations lead to the same entropy production
$\sigma$ per unit
volume, as it should.}
\be
(c=2) \qquad \qquad \sigma_{\rm irr} = - \left( 1 +
\frac{m_1n_1}{m_2n_2}\right) \; \vec{J}_1 \cdot \vec{\nabla}\frac{\mu_1}{T}
\; .
\label{entr.prod.mutual.diff}
\ee
If the tracer particles of mass $m_1$ are much lighter than the other
particles of mass $m_2$ and, moreover, if the tracer particles are dilute
\be
m_1 \, n_1 \ll m_2 \, n_2 \; ,
\ee
the entropy production reduces to the simple expression
\be
\sigma_{\rm irr} = -  \vec{J}_1 \cdot \vec{\nabla}\frac{\mu_1}{T}
\; ,
\ee
corresponding to the simple diffusion of the tracer particles.
The heavy particles fix the center of mass with respect to
which the transport is defined and they play the role of scatterers
for the lighter particles. The thermodynamic flux of
this diffusive process can thus
be identified as the current of the tracer particles
\be
\vec{J}^{\rm (D)} \equiv \vec{J}_1 \; ,
\ee
while the associated thermodynamic force is given in terms of the gradient of
the chemical potential of the tracer particles as \cite{Ni79}
\be
\vec{X}^{\rm (D)} = -\vec{\nabla}\frac{\mu_1}{T} \; .
\ee
Moreover, for dilute tracer particles,
the chemical potential depends logarithmically on the tracer
density $n_1$ according to
\be
\frac{\mu_1}{T} = k_{\rm B} \; \ln \frac{n_1}{n_1^0} \; ,
\label{mu1}
\ee
where $n_1^0$ is a reference density \cite{BeRiRo80}.
Hence, in a dilute system where the temperature, $T$, is
constant, the thermodynamic force becomes
\be
\vec{X}^{\rm (D)} = -k_{\rm B}\; \frac{\vec{\nabla}n_1}{n_1} \; ,
\label{thermoforce}
\ee
which is independent of the temperature.  The thermodynamic
force is simply given in terms of the
gradient of tracer density, and drives the
irreversible process of matter exchange at the
macroscopic level.
This irreversible process occurs even if the tracer
particles are so dilute that they
never meet each other.  The point is that
the thermodynamic force of diffusion is a concept associated with
a collection of particles, in contrast
to the microscopic forces which directly act
on the particles themselves.  As we show in Sec. \ref{sec-Hthm},
such a force is indeed manifested as a Newtonian force
at the level of momentum balance of the light particles
but the type of balance one refers to here pertains to the
{\it average} momentum taken over the probability
distribution of the particles, as usual in fluid mechanics
and in macroscopic physics in general.
It may be worth
recalling that in some molecular dynamics-inspired
mechanisms of thermostating designed to sustain
nonequilibrium steady states with shear or heat flow, the nonequilibrium
constraint (an analog of the local thermodynamic
force considered here) does modify the dynamics of the
individual particles \cite{EvMo90,Ho99}.
In our view, this ingenious scheme,
which stimulated many developments, is to be looked at as
a prototypical model along with other thermostating
mechanisms \cite{LeBe57,Le59,ChLe95,ChLe97,EcPiRe99,KlRaNi00}.
In this connection we mention that nonequilibrium steady states with
processes of matter exchange would require nonequilibrium constraints
on particle numbers, that is,
chemiostating mechanisms.

In the linear range of irreversible processes,
the flux is proportional to
the thermodynamic force. We shall write this proportionality
in a form displaying Fick's diffusion coefficient $\cal D$
rather than Onsager phenomenological coefficient,
\be
\vec{J}^{\rm (D)} = - {\cal D}\; \vec{\nabla}n_1 \; ,
\ee
since $\cal D$ varies weakly in a wide range of values of $n_1$.
Then the contribution of diffusion to the irreversible entropy
production is given by
\be
\sigma_{\rm irr} = \vec{J}^{\rm (D)}\cdot \vec{X}^{\rm (D)} = k_{\rm B}\;
{\cal D} \; \frac{(\vec{\nabla} n_1)^2}{n_1} \ge 0 \; .
\label{irr-entrprod}
\ee
The diffusion
coefficient $\cal D$ can be positive and finite even in the case of
dilute tracer particles in a fluid.  Since each tracer
particle undergoes a random walk due to collisions with the
surrounding fluid particles, a statistical collection of them appear
to be driven from the regions where they are more
concentrated toward the regions where their concentration is lower.
The statistical character of the random walk is at the origin of
the irreversible production of entropy.
Thus, the absence of mutual interaction between the
tracer particles does not prevent the existence of a thermodynamic force
contributing to entropy production.

In conclusion, a careful application of
the notion of thermodynamic force in the
case of diffusion, as explained above,
shows that the Cohen-Rondoni argument
that, in the case of diffusion, the thermodynamic force should
act on the particles {\it themselves} \cite{CoRo02}
is incorrect.  It results from a failure to appreciate
the fact that the thermodynamic force of
diffusion has a statistical origin and should
not be conceived as a microscopic
force which acts on the particles themselves.
Indeed, the tracer particles can be so
dilute that they never come within
interaction range among themselves although
the thermodynamic force of diffusion is
still non-vanishing and contributing to
entropy production.  Moreover,
the derivation here above shows that the
entropy production does not vanish
in the Lorentz-gas limit $(m_1/m_2)\to 0$
where the heavy particles are
immobile.

In the next section, we show starting
from a microscopic description that diffusion
in the Lorentz gas is indeed well described by
irreversible thermodynamics, and that the
entropy produced corresponds to
physical work that might be done on a
moving piston, for example.


\section{The irreversible thermodynamics of the Lorentz gas}
\label{Lorentz}

In this section, we show, contrary to the claims by Cohen and
Rondoni \cite{CoRo02}, that the Lorentz gas can approach an equilibrium state
through a local equilibrium and that a thermodynamic force as well
as a friction force do exist in the Lorentz gas.
Furthermore, we show that diffusion in the Lorentz gas
yields a positive entropy production
and that such an entropy production has a clear
physical implication.

\subsection{$H$-theorem for the random Lorentz gas}
\label{sec-Hthm}

The Lorentz gas model was introduced by Lorentz in
1905 as a model for transport of electrons in a solid \cite{Lo05}. It
consists of a collection of fixed scatterers, and a collection of
moving particles which do not interact with each other, but only with
the fixed scatterers according to a well-defined potential energy
function. Today, the Lorentz gas is often used as a model of diffusion of
light particles among heavy ones \cite{vBe82,De00,CC70,Ha74}.  The heavy
particles are
immobile on the time scale of motion of the light
particles, and the light particles are considered as not interacting
with each other.  At each elastic collision between light and heavy particles,
energy is nearly conserved although
momentum is not.  Therefore, the randomization of the velocity direction is
very fast albeit the randomization
of energy occurs on a much longer time scale.  This decoupling of time
scales does not preclude the existence of
a well-defined diffusion coefficient on each energy shell.

To set the stage we first analyze the transport properties
and the thermodynamics of the Lorentz gas from the standpoint
of kinetic theory.  To this end we consider the random Lorentz gas
with dilute scatterers, a version of the Lorentz model in which
the diffusion coefficient has been evaluated some time ago
using kinetic theoretic methods \cite{vBe82,Ha74}.  More specifically,
we consider a two-dimensional dilute random Lorentz gas.
The scatterers are assumed to be hard disks of radius $a$ which are
uniformly distributed on the plane with the density $n_2$.
The motion of tracer particles is
described by a linear Boltzmann equation,
also known as the Boltzmann-Lorentz equation \cite{vBe82,BoBuSi83}
\be
\partial_t \; f + \vec{v}\cdot \vec{\nabla} f = \frac{n_2av}{2}
\int_{-\pi}^{+\pi}
d\varphi' \left\vert\sin\frac{\varphi-\varphi'}{2}\right\vert \left[
f(\vec{r},\varphi')-f(\vec{r},\varphi)\right] \; ,
\label{BLeq}
\ee
for the number density $f(\vec{r},\varphi)$ of tracer particles at position
$\vec{r}$ with
velocity angle $\varphi$, where $\vec{v}=v(\cos\varphi,\sin\varphi)$ is the
velocity
of the particles, $v$ their speed which is conserved as well as
their kinetic energy $\epsilon=\frac{1}{2}m_1v^2$.

The Boltzmann-Lorentz equation (\ref{BLeq}) obeys a $H$-theorem based on the
local entropy density
\be
s(\vec{r}) = k_{\rm B} \int_{-\pi}^{+\pi} d\varphi \; f(\vec{r},\varphi) \;
\ln \frac{{\rm e}\;  f^0}{f(\vec{r},\varphi)} \ ,
\ee
where $f^0$ is a constant, and ${\rm e}=2.718\ldots$ denotes the Naperian base.
Indeed, the balance equation for this entropy density is given by
\be
\partial_t\; s = -\vec{\nabla}\cdot\vec{J}_s + \sigma_{\rm irr} \; ,
\label{bal.entropy}
\ee
with the entropy current
\be
\vec{J}_s = k_{\rm B} \int_{-\pi}^{+\pi} d\varphi \; \vec{v} \;
f(\vec{r},\varphi) \;
\ln \frac{f^0}{f(\vec{r},\varphi)} \ ,
\ee
and the local entropy production
\be
\sigma_{\rm irr} = \frac{1}{4}\;  k_{\rm B} \, n_2 \, a \, v  \int d\varphi \;
d\varphi' \;
\left\vert\sin\frac{\varphi-\varphi'}{2}\right\vert \left[
f(\vec{r},\varphi')-f(\vec{r},\varphi)\right] \;
\ln\frac{f(\vec{r},\varphi')}{f(\vec{r},\varphi)} \geq 0 \ .
\label{Hthm}
\ee
We emphasize that this $H$-theorem holds although all particles
have the same energy $\epsilon=\frac{1}{2}m_1v^2$.
The mechanism at the origin of
this $H$-theorem is the randomization of the velocity angle,
which establishes a
local equilibrium in velocity direction.  We notice that
the existence of such a local equilibrium has already been
pointed out by Lebowitz and Spohn in their work on heat conduction
in the random Lorentz gas \cite{LeSp78}.

One can expand the density function as a Fourier series in the velocity angle
$\varphi$ as
\be
f = \sum_{k=-\infty}^{+\infty} f_k \; {\rm e}^{{\rm i}k \varphi} \; .
\ee
As a consequence of the Boltzmann-Lorentz equation (\ref{BLeq}) the Fourier
components $f_k$ obey the following coupled differential equations
\be
\partial_t \; f_k + v \; \frac{\partial_x-{\rm i}\partial_y}{2} \; f_{k-1}
+ v \; \frac{\partial_x+{\rm i}\partial_y}{2} \; f_{k+1} =
\frac{8k^2}{1-4k^2} \, n_2
\, a \, v \; f_k \; .
\ee
For particle distributions close to the microcanonical equilibrium, we keep
only the first Fourier components so that
\be
f(\vec{r},\varphi) \simeq \frac{1}{2\pi} \left[ n_1(\vec{r}) + \frac{2}{v^2}
\; \vec{v}\cdot\vec{J}_1(\vec{r}) \right] \ ,
\label{expand}
\ee
where the density of tracer particles and their current
are given by
\begin{eqnarray}
n_1(\vec{r}) &=& \int_{-\pi}^{+\pi} f(\vec{r},\varphi) \; d\varphi \; , \\
\vec{J}_1(\vec{r}) &=& \int_{-\pi}^{+\pi} \vec{v} \; f(\vec{r},\varphi) \;
d\varphi
\; .
\end{eqnarray}

Substituting expansion (\ref{expand}) in the Boltzmann-Lorentz
equation (\ref{BLeq}) and retaining the dominant terms,
we get the balance equations
for the number of tracer particles and for their current as
\begin{eqnarray}
\partial_t \; n_1 &=& - \vec{\nabla}\cdot \vec{J}_1 \; ,
\label{local.cons.nber}\\
\partial_t \; \vec{J}_1 &\simeq&
-\frac{8}{3} \, n_2 \, a \, v \; \vec{J}_1
- \frac{v^2}{2} \vec{\nabla} n_1 \; .
\label{bal.current}
\end{eqnarray}
Over a time scale longer than the relaxation time $\tau=\frac{3}{8n_2av}$, the
current is driven by the gradient of density and we obtain
from Eq. (\ref{bal.current}) Fick's law on
the energy shell as
\be
\vec{J}_1 \simeq - {\cal D} \; \vec{\nabla} n_1 \; .
\ee
After substituting in Eq. (\ref{local.cons.nber}) we obtain the diffusion
equation
\be
\partial_t n_1(\vrr) = {\cal D}\; \nabla^2 n_1(\vrr) \; ,
\label{diff0}
\ee
with the energy-dependent diffusion coefficient
\be
{\cal D} = \frac{3 \, v}{16 \, a \, n_2} \; ,
\label{diff-rndm}
\ee
where $v=\sqrt{2\epsilon/m_1}$ is the speed of particles.  Under these
conditions,
the entropy production (\ref{Hthm}) is straightforwardly shown to reduce to
\be
\sigma_{\rm irr} \simeq k_{\rm B} \; \frac{\vec{J}_1^2}{{\cal D}\, n_1}
\simeq k_{\rm B}\; {\cal D} \; \frac{(\vec{\nabla} n_1)^2}{n_1}
\geq 0 \; ,
\label{entrprod0-rndm}
\ee
which is the expression expected from irreversible thermodynamics,
here shown to hold on a single energy shell.

We next show that the thermodynamic force of diffusion
(\ref{thermoforce}) appears explicitly in the balance equation for the
momentum of the gas of tracer particles.  We notice that the momentum density
$\vec{g}_1$ of the
tracer particles is directly proportional to the particle current $\vec{J}_1$
according to
\be
\vec{g}_1 \equiv m_1 \; \vec{J}_1 \; .
\ee
The balance equation (\ref{bal.current}) can thus be interpreted as the
balance equation for the momentum of the gas of tracer particles. Moreover,
we can
introduce the partial mass density of the tracer particles as
$\rho_1=m_1n_1$ and their drift velocity as
\be
\vec{u}_1 \equiv \frac{\vec{g}_1}{\rho_1} = \frac{\vec{J}_1}{n_1} \; .
\ee
This drift velocity can be considered as the velocity of an element of the
gas of
tracer particles.  Since Eq. (\ref{bal.current}) gives the balance equation for
momentum, we can now derive the time evolution of an element of the tracer
gas as
\be
\rho_1 \; \frac{d\vec{u}_1}{dt} \simeq
\underbrace{- \beta \; \rho_1
\; \vec{u}_1}_{\vec{F}_{\rm friction}}
\underbrace{-\epsilon \; \vec{\nabla}
n_1}_{\vec{F}_{\rm thermo.}}
\label{Newton}
\ee
where we have neglected terms which are cubic in the gradients of the density.
We notice that, to the order we keep, neglecting the cubic terms
allows us to replace the partial time derivative in (\ref{bal.current})
by a total time derivative in (\ref{Newton}).
The constant $\beta=\frac{8}{3}n_2av$ is the friction coefficient, while
$\epsilon$
is the kinetic energy of the particles. Eq. (\ref{Newton}) can be
interpreted as
Newton's equation for the element of the tracer gas.  The left-hand side
contains
the mass of the gas element multiplied by its acceleration while the different
Newtonian forces (per unit volume) acting on this
gas element can be identified in the two terms of the right-hand side.

The first term is the Newtonian force of friction.
The presence of this term
contradicts the claim by Cohen and Rondoni that
``fluxes of particles are moving {\it frictionless}"
in the Lorentz gas (Ref. \cite{CoRo02}, p. 121).
On the contrary, the infinitely heavy particles which are
the scatterers produce a friction on the tracer gas because of the interaction
of each tracer particle with the scatterers.
We notice that conservation of total energy is here reduced to the
property that the microscopic kinetic energy of the
individual particles, $e=\epsilon \, n_1$, is locally conserved according to
\be
\partial_t\; e + \vec{\nabla}\cdot (e \; \vec{u}_1)=0 \; .
\ee
Friction terms will be manifested when $e$ is decomposed
into a macroscopic kinetic energy part, $n_1\vec{u}_1^2/2$, and
an internal energy part.  This is exactly along the same lines
as the derivation of the balance equations of
macroscopic physics.

The second term in the right-hand side of Eq. (\ref{Newton})
involves the gradient of the density and can thus be
identified as the Newtonian force driven by the thermodynamic force of
diffusion
(\ref{thermoforce}).  Indeed, we have a proportionality between both given by
\be
\vec{F}_{\rm thermo.} = \frac{n_1\, \epsilon}{k_{\rm B}} \; \vec{X}^{\rm
(D)} \; .
\label{thermoNewton}
\ee
This result shows that, in the Lorentz gas, the thermodynamic force acts as
a ``real force" albeit on the statistical collection
of particles which compose the gas element,
even if the particles have no mutual interaction and, furthermore, conserve
their energy during their whole motion.  This is very similar to the role
of pressure gradient in the momentum balance of a volume element
of a fluid, which
is present even in the limit of noninteracting particles.
It is clear that this thermodynamic force does not act on the
individual particles themselves.  The concept of thermodynamic
force applies to the statistical level of description but not to the
microscopic one.

In the stationary state, the two terms in the right-hand side of Eq.
(\ref{Newton})
cancel, and one has
\be
\vec{u}_1 \simeq - \frac{\epsilon}{\beta \, \rho_1} \; \vec{\nabla} n_1 \; .
\ee
The entropy production (\ref{entrprod0-rndm}) of diffusion can then be
expressed in
terms of the dissipation arising from the friction
\be
\sigma_{\rm irr} \simeq k_{\rm B} \; \frac{\beta \,
\rho_1}{\epsilon}  \; \vec{u}_1^{\, 2} \; .
\label{entrprod0-rndm-friction}
\ee

In Ref. \cite{CoRo02}, Cohen and Rondoni claim that there is no
thermodynamic force for diffusion in the Lorentz gas, because there is
no local equilibrium state described by a Maxwell-Boltzmann
distribution with space and time dependent macroscopic quantities -
the local density and temperature, and the local mean velocity. Here, we have
demonstrated that despite its simplicity, the random Lorentz gas has all the
ingredients needed for a description of its transport processes by
irreversible thermodynamics. Thus the claims to the contrary by these
authors appear to be mistaken. In our opinion they have not made a
sufficiently careful distinction between the microscopic and
macroscopic levels of description of the system, and they have not
appreciated the fact that the approach to equilibrium in a Lorentz gas
with a small density gradient also proceeds through a state of local
equilibrium.
Here the local equilibrium is not a state of near-Maxwellian velocity
distribution
but, rather, a state which is nearly uniform in the velocity direction,
with a local
density, even though all of the moving particles have the same energy.
This local equilibrium is at the very origin of
diffusion itself and, thus, of irreversible entropy production on each
energy shell.

Finally, we notice that the random Lorentz gas is a useful model for the
electronic conductivity at low temperature in a solid
with impurities, and leads to a determination of the so-called
residual resistivity.  In this case, it is known that the conductivity
is related to the diffusion coefficient and that both should be evaluated
on the energy shell corresponding to the Fermi energy of the electron gas
\cite{BeKi94}.  Diffusion as well as entropy production
on a single energy shell have
therefore experimental relevance.

\subsection{The periodic Lorentz gases as models of diffusion}

We now proceed to the extension of the previous
refutation of the criticisms by Cohen and Rondoni
to the case of the periodic Lorentz gases
where diffusion was rigorously proved.
Using the methods of dynamical systems theory, which are free of
the statistical hypotheses underlying kinetic theory,
we show in the present subsection
that, for these Lorentz gases, local equilibrium in the form discussed
in Sec. \ref{sec-Hthm} and
Fick's law also hold on each energy shell and,
in the following subsection,
that this results in a positive entropy production.

In 1980, Bunimovich and Sinai proved that the
two-dimensional hard-disk periodic Lorentz gas has a positive and finite
diffusion coefficient if the horizon is
finite, i.e., if no trajectory exists running through the lattice of hard
disks without elastic collision
\cite{BuSi80}.  A similar result has been proved by Knauf for the
two-dimensional periodic Lorentz gas composed of
scatterers with Yukawa potentials under the condition that the energy of
the moving particle is large enough
\cite{Kn87}.  In addition, these Lorentz gases have been proved to be ergodic
and mixing. The mixing property holds on each energy shell,
and leads to the establishment of a {\it
local equilibrium in velocity direction}. As in the case of the
random Lorentz gas discussed above, this local equilibrium may
be described by a spatial and velocity distribution which is nearly uniform
in the
directions of the velocity of the moving particles, and the spatial density
varies only over distances large compared to the mean free path of the
particles.
This local equilibrium is sufficient for a transport
by diffusion to occur, compatible with the laws of irreversible
thermodynamics.

Mathematical proofs of the existence of diffusion in periodic
Lorentz gases given by Bunimovich and Sinai and Knauf \cite{BuSi80,Kn87},
as well as extensions to higher orders
in density gradients by Chernov and Dettmann
\cite{ChDe00}, all imply the existence of a well behaved
diffusion equation for these models.
That is, on large spatial scales, compared to the mean free path of a particle,
the diffusion occurs on each energy shell so that
the density $f(\vrr,\epsilon)$ of light particles with energy
$\epsilon$ at position $\vrr$ obeys the diffusion equation
\be
\partial_t \, f(\vrr,\epsilon)\simeq {\cal D}(\epsilon) \ \nabla^2
f(\vrr,\epsilon) \; ,
\label{diff-p}
\ee
where ${\cal D}(\epsilon)$ is the diffusion coefficient on the shell of
energy $\epsilon$, as proved in Ref. \cite{BuSi80}.

The diffusion equation (\ref{diff-p}) can be rewritten in the form of a local
conservation law for the density $f(\vrr,\epsilon)$ as
\be
\partial_t \, f\ + \ \vec{\nabla}\cdot
\vec{J}_f = 0 \; ,
\label{conserv}
\ee
with the current
\be
\vec{J}_f \simeq -{\cal D}(\epsilon) \; \vec{\nabla} f(\vec{r},\epsilon) \; .
\label{current-f}
\ee
This result, which was directly derived
from the microscopic dynamics
in Ref. \cite{Ga96}, shows that Fick's law holds
on each energy shell in the Lorentz gas.
This is a consequence of the property
of local equilibrium in the velocity direction
on each energy shell, and responds to the doubts expressed by Cohen and Rondoni
\cite{CoRo02} about the existence of a local equilibrium state
and a corresponding Fick's law
in the periodic Lorentz gases.

At each elastic collision the energy $\epsilon$ of the particle is conserved
so that the particle keeps its energy.  Particles with different energies can
thus be distinguished.  Hence, the gas can be envisaged as a mixture
of different gases with particles having different energies.  These different
gases ignore each other and each of them is the stage of
an irreversible process of diffusion according to Eq. (\ref{diff-p}).
In such a gas, the number of light particles per unit volume around the
spatial point $\vrr$ is defined as
\be
n_1(\vec{r}) \equiv \int_0^{\infty} f(\vec{r},\epsilon) \;
d\epsilon \; .
\label{n1}
\ee
We can consider separately the gases of particles
with different energies.  If we remove from the system
all the particles except those having their
energy between $\epsilon$ and $\epsilon+\Delta\epsilon$
(with $\Delta\epsilon$ arbitrarily small),
the particle density is given by
\be
n_1(\vec{r}) \simeq f(\vec{r},\epsilon) \; \Delta\epsilon \; .
\label{n1-p}
\ee
As a consequence of the diffusion equation (\ref{diff-p}),
the density (\ref{n1-p}) obeys the diffusion equation
\be
\partial_t \, n_1(\vrr) = {\cal D}\; \nabla^2 n_1(\vrr) \; ,
\label{diff}
\ee
with the diffusion coefficient ${\cal D}={\cal D}(\epsilon)$.
We notice that Eq. (\ref{diff}) as well as Eq. (\ref{diff-p})
are also found in the random Lorentz gas with the diffusion
coefficient (\ref{diff-rndm}) according to the kinetic theory
of the previous subsection.

The diffusion equation (\ref{diff}) describes processes in which the initial
state is created by a spatial disturbance in the particle density $n_1$.
Thereafter, the system relaxes back to a spatially uniform
state according to Eq. (\ref{diff}).  In this spatially uniform state,
all the particles still have the same energy $\epsilon$
while their velocity angle is uniformly distributed in the interval $[0,2\pi[$
and their position is uniformly distributed in the whole space accessible
to the particles.  This uniform state is a microcanonical state for each
one of the particles.  This state plays the role of the state of
thermodynamic equilibrium in the Lorentz gas.

\subsection{The thermodynamics of the Lorentz gas}

Here, we show by using the methods of thermodynamics
that diffusion in the Lorentz gases has a positive entropy production
on each energy shell.  We first establish the
equilibrium thermodynamics of the Lorentz gas
in the microcanonical ensemble
and subsequently derive the balance equation for the entropy density
in order to obtain the local entropy production.

The equilibrium thermodynamics of the hard-ball Lorentz gas is established
as follows.  Let us consider a
gas of $N_1$ light particles without mutual interaction
and moving in a spatial domain
$\cal Q$ bounded internally by the hard balls and externally by a
wall enclosing the Lorentz gas in a square, say.
The volume of the domain
$\cal Q$ is ${\cal V}={\rm Vol}({\cal Q})$.
We notice that this volume is related to
the physical volume $V$ by
\be
{\cal V} = V - N_2 \; v_2 \; ,
\ee
where $N_2$ is the number of hard balls and $v_2$ is the volume of one hard
ball. The phase space of this gas is
\be
{\cal M}=\left\{ (\vrr_1,\vp_1,...,\vrr_{N_1},\vp_{N_1})\vert
\vrr_j\in{\cal Q},
\vp_j\in\mbox{\mb R}^d, x_1<x_2<\cdots
<x_{N_1}\right\} \; ,
\ee
where $\vrr_j=(x_j,y_j,z_j)$ and
the last conditions provide the ordering of indices required in the
case of identical particles.  The phase-space volume element is
\be
d\G = d\vrr_1\; \cdots \; d\vrr_{N_1} \; d\vp_1 \; \cdots \; d\vp_{N_1} \; .
\ee
The system being ergodic, the equilibrium invariant distribution is given
by a microcanonical distribution for a system of $N_1$ independent particles
of kinetic energy between $\epsilon$ and $\epsilon+\Delta\epsilon$
(with $\Delta\epsilon$ arbitrarily small):
\be
F(\G) = \cases{ \frac{\displaystyle N_1!}{\displaystyle{\cal V}^{N_1}} \;
\left[ \frac{
\displaystyle \Gamma\left(\frac{d}{2}\right) \; \epsilon }
{\displaystyle \left(2\pi m_1 \epsilon\right)^{\frac{d}{2}} \;
\Delta\epsilon}
\right]^{\displaystyle N_1}
\qquad {\rm for} \quad \displaystyle\epsilon <
\frac{\vec{p}_j^{\, 2}}{2m_1}
< \epsilon+\Delta\epsilon \; , \cr\cr
0 \qquad {\rm otherwise} \; ,\cr}
\label{distrib}
\ee
which is normalized according to
\be
\int_{\cal M} F(\G) \; d\G = 1 \; .
\ee

The entropy can be calculated as Gibbs' coarse-grained entropy \cite{Gi1902}
\be
S=-k_{\rm B} \sum_i p_i \ln p_i \; ,
\label{Gibbs}
\ee
by coarse graining the phase space into cells of size $\Delta\G =
\Delta^{N_1d}r\; \Delta^{N_1d}p=
(\Delta r\Delta p)^{N_1d}$.  This is the microscopic definition of the
entropy which has been adopted in Refs. \cite{GiDoGa00,DoGaGi02}
for nonequilibrium distribution.
The probability for the system to belong to the
$i^{\rm th}$ cell is
\be
p_i = \int_{i^{\rm th} \; {\rm cell}} F \; d\G \simeq F_i \; \Delta\G \; ,
\ee
so that
\be
S\simeq k_{\rm B} \left\langle \ln \frac{1}{F(\G)\; \Delta\G}\right\rangle
\; .
\label{Gibbs.entropy}
\ee
A straightforward calculation of the entropy (\ref{Gibbs.entropy}) for the
equilibrium distribution (\ref{distrib}) leads to
the formula
\be
S \simeq k_{\rm B} \; N_1 \; \ln \frac{{\cal V} \, {\rm e} \,
(2\pi m_1 \epsilon)^{\frac{d}{2}} \, \Delta\epsilon}{N_1 \
\Gamma\left(\frac{d}{2}\right) \, \epsilon \, (\Delta r\Delta p)^{d}} \; ,
\label{entropy-1}
\ee
in the limit $N_1,{\cal V}\to\infty$, keeping the ratio $N_1/{\cal V}$
constant (where, again, ${\rm e}=2.718\ldots$ denotes the Naperian base).  The energy
of the gas is
\be
E = N_1 \; \epsilon \; .
\label{energy}
\ee
We notice that $S$ is a well-defined differentiable function.
As a consequence, differentiating both the
formula (\ref{entropy-1}) for entropy and the energy
equation of state (\ref{energy}) with respect to the variables
$V$, $N_1$, $N_2$, and $\epsilon$, we
obtain the following identity which is known as the Gibbs relation
\cite{dGrMa62}
\be
dS = \frac{1}{\cal T}\; dE + \frac{P}{\cal T} \; dV - \frac{\mu_1}{\cal T}
\; dN_1 -
\frac{\mu_2}{\cal T} \; dN_2 \; ,
\label{Gibbs.relation}
\ee
with the inverse temperature
\be
\frac{1}{\cal T} =  \frac{d-2}{2} \; \frac{k_{\rm B}}{\epsilon} \ ,
\label{itemperature}
\ee
the equation of state
\be
\frac{P}{\cal T} = \frac{N_1 \; k_{\rm B}}{V-N_2\, v_2} \; ,
\ee
and the chemical potentials given by
\be
\frac{\mu_1}{\cal T} = k_{\rm B} \; \ln  \frac{N_1 \; {\rm e}^{\frac{d}{2}-1}
\; \Gamma\left(\frac{d}{2}\right) \, \epsilon \; (\Delta r \Delta p)^d}
{(V-N_2\, v_2) \, (2\pi m_1 \epsilon)^{\frac{d}{2}} \, \Delta\epsilon} \; ,
\label{67}
\ee
for the light particles and by
\be
\frac{\mu_2}{\cal T} = \frac{N_1 k_{\rm B} v_2}{V-N_2\, v_2} \; ,
\label{68}
\ee
for the hard balls.
It follows from Eqs. (\ref{67}) and (\ref{68}) that
$\mu_1/{\cal T}$ and $\mu_2/{\cal T}$,
which are related to the thermodynamic force
of diffusion, are well defined. Notice that the inverse temperature
happens to vanish in the two-dimensional case
[cf. Eq. (\ref{itemperature})], while the ratios of pressure and
chemical potential to temperature are non-zero.

If we introduce the intensive quantities defined by Eq. (\ref{densities}),
we obtain the other Gibbs relation
\be
ds = \frac{1}{\cal T} \; de - \frac{\mu_1}{\cal T} \; dn_1
- \frac{\mu_2}{\cal T} \; dn_2 \; ,
\label{local.Gibbs.relation}
\ee
where the entropy density is given by
\be
s = k_{\rm B} \; n_1 \; \ln
\frac{{\rm e}^{\frac{d}{2}} \; n_1^0}{n_1} \; ,
\label{entropy.density}
\ee
with the reference density
\be
n_1^0 =  \frac{(1-n_2v_2) \,
(2\pi m_1 \epsilon)^{\frac{d}{2}} \,
\Delta\epsilon}{ {\rm e}^{\frac{d}{2}-1} \;
\Gamma\left(\frac{d}{2}\right) \, \epsilon \,
(\Delta r\Delta p)^{d}} \; .
\label{ref.density}
\ee
In Eq. (\ref{local.Gibbs.relation}),
the energy density is given by
\be
e = n_1 \; \epsilon \; ,
\label{energy.density}
\ee
the light-particle chemical potential  by
\be
\frac{\mu_1}{\cal T} = k_{\rm B} \; \ln  \frac{n_1}{n_1^0} \; ,
\ee
the hard-ball chemical potential by
\be
\frac{\mu_2}{\cal T} = \frac{n_1 k_{\rm B} v_2}{1-n_2\, v_2} \; ,
\ee
and the inverse temperature by Eq. (\ref{itemperature}).

The relation (\ref{local.Gibbs.relation}) holds for the system at equilibrium.
Its local equilibrium analog would amount to
a supposition that the density $n_1$ of light particles presents small
variations over large spatial scales while the energy $\epsilon$ and the
hard-ball density $n_2$ remain constant, and
the entropy density $s$ varies in such a way that
the Gibbs relation (\ref{local.Gibbs.relation})
remains satisfied locally.
The relaxation to such a state of local
equilibrium finds its full justification in the complete
{\it ab initio} derivation of entropy
production given in Ref. \cite{DoGaGi02}.
The condition of validity is that the randomization
of the velocity direction is faster than the relaxation
of spatial inhomogeneities of
the particle densities on each energy shell.
The randomization time is given by the intercollisional
time $\tau_{\rm coll.} \sim \ell/v$
which is estimated by the ratio of the mean free path $\ell$
to the speed $v=\sqrt{2\epsilon/m_1}$.  The relaxation time is given by
the diffusion time $t_{\rm diff.}\sim L^2/{\cal D}$
where $L$ is the wavelength characterizing the spatial
inhomogeneities.  Since ${\cal D}\sim v \ell$,
the condition $t_{\rm diff.} \gg \tau_{\rm coll.}$
thus requires that the wavelength
of the particle density is larger than the mean free path:
$L \gg \ell$.  This condition is independent of the
speed so that it holds uniformly on all the energy shells.

For a nonequilibrium situation with a local variation of
the density $n_1$ of light particles,
we can now obtain the balance equation for the entropy density
(\ref{entropy.density}) on the basis of the knowledge that the particle
density $n_1$ obeys the diffusion equation (\ref{diff}) in the Lorentz
gas.  The density $n_2$ of hard balls is constant in time and uniform in
space.   The entropy balance equation is obtained by taking the time
derivative of the entropy density
(\ref{entropy.density}), replacing $\partial_tn_1$
by the right-hand side of Eq. (\ref{diff}),
and splitting the result into a divergence term and a rest term
which can then be identified as the local source of entropy,
i.e. the irreversible entropy production.  Performing this calculation
yields the same balance equation as (\ref{bal.entropy}) but here
with the entropy current
\be
\vec{J}_s = -{\cal D}\; k_{\rm B} \; \left(\ln\frac{{\rm e}^{\frac{d}{2}-1}\,
n_1^0}{n_1}\right) \; \vec{\nabla} n_1 = -{\cal D}\; \vec{\nabla} s \; ,
\ee
and the local irreversible entropy production
\be
\sigma_{\rm irr} = k_{\rm B}\; {\cal D} \; \frac{(\vec{\nabla} n_1)^2}{n_1}
\geq 0 \; ,
\label{entrprod0}
\ee
which is always nonnegative by the Second Law of thermodynamics.
The conclusion is here that the standard thermodynamic reasoning
to derive entropy production \cite{dGrMa62}
can be developed in the Lorentz gas
and yields the same result as Eq. (\ref{entrprod0-rndm})
obtained by kinetic theory in the case of the random Lorentz gas.
The present derivation is more general
in the sense that it also applies to the periodic
Lorentz gas.

We now consider a gas of tracer particles moving in the same
system as above, though with a nontrivial distribution of energy
$f(\vec{r},\epsilon)$ obeying the diffusion equation
(\ref{diff-p}).  As mentioned above, such a gas can be considered
as a mixture of different gases with particles having
their energy defined within intervals $\Delta\epsilon$.
The local entropy density of this gas is given by the sum of the local
entropies (\ref{entropy.density})-(\ref{ref.density})
for the particles with different energies.  Since
the density of the particles of energy $\epsilon$ is given by
$n_1(\vec{r}) \simeq f(\vec{r},\epsilon) \Delta\epsilon$,
we get the local entropy density of
the whole gas as
\be
s(\vec{r}) =  \int_0^{\infty} d\epsilon \; k_{\rm B} \; f(\vec{r},\epsilon)
\; \ln \frac{(1-n_2v_2) \, {\rm e} \,
(2\pi m_1 \epsilon)^{\frac{d}{2}} }{f(\vec{r},\epsilon) \
\Gamma\left(\frac{d}{2}\right) \, \epsilon \, (\Delta r\Delta p)^{d}} \; .
\label{local.entropy.density-p}
\ee
By performing a similar calculation as above but using the diffusion equation
(\ref{diff-p}), we can obtain a balance equation similar to Eq.
(\ref{bal.entropy}) now for the full entropy
(\ref{local.entropy.density-p}).  In this case, the local
irreversible entropy production in the Lorentz gas is given by
\be
\sigma_{\rm irr}(\vec{r}) = \int_0^{\infty} d\epsilon \;
k_{\rm B}\; {\cal D}(\epsilon) \;
\frac{\left[\vec{\nabla} f(\vec{r},\epsilon)\right]^2}{f(\vec{r},\epsilon)}
\geq 0 \; .
\label{entrprod-p}
\ee
which is one of the central results of this paper.
This result shows that the Lorentz gas with a distribution
of energy can be considered as a mixture of different gases
distinguished by the energy of their particles and that
diffusion on each energy shell of the gas separately
contributes to a positive entropy production.
The main point is that, already in the microcanonical ensemble,
the entropy production is positive with the typical form (\ref{entrprod0})
expected from irreversible thermodynamics.  The origin of this
entropy production holds in the randomization
of the velocity angle due to mixing in the Lorentz gas,
which leads to the local equilibrium in the velocity direction.

In Refs. \cite{TaGa99,TaGa00,Ga97b,GiDoGa00,DoGaGi02},
it was shown that a similar calculation can be carried out {\it ab
initio} at the microscopic level of description, which leads to the very same
expression (\ref{entrprod0}) that we have here derived using
the local Gibbs relation (\ref{local.Gibbs.relation}) and the diffusion
equation (\ref{diff-p}).

In conclusion, equations (\ref{entrprod0-rndm}),
(\ref{entrprod0}) or (\ref{entrprod-p}) clearly
demonstrate that entropy production does not vanish in the Lorentz
gas, contrary to the claim by Cohen and Rondoni in Ref. \cite{CoRo02} p. 127
that there is no entropy production in the Lorentz gas.

We notice that our result (\ref{entrprod-p})
applies not only to the Lorentz gas in which
the mass ratio $m_2/m_1$ is infinite
but also to a binary mixture where the mass
ratio $m_2/m_1$ is large but finite.
In this latter case,
there still exists a mechanism which
thermalizes and drives the velocity distribution toward
the Maxwell-Boltzmann distribution even
if this mechanism is slow.
Under this assumption, the local distribution
of energy can be supposed to be factorized
into a local particle density $n_1(\vrr)$
multiplied by a uniform equilibrium Boltzmann energy
distribution of given temperature
$T$ as
\be
f(\vrr,\epsilon) \simeq n_1(\vrr) \; \frac{\eta(\epsilon)\, {\rm
e}^{-\epsilon/k_{\rm
B} {T}}}{\int_0^{\infty}\eta(\epsilon)\,{\rm e}^{-\epsilon/k_{\rm B}
{T}}\; d\epsilon} \; ,
\label{distrib-macro}
\ee
where $\eta({\epsilon})$ is the density of states at energy $\epsilon$.
Consequently, the local entropy density
(\ref{local.entropy.density-p}) takes the form
\be
s(\vec{r}) =  k_{\rm B} \; n_1(\vec{r})
\; \ln \frac{(1-n_2v_2) \, {\rm e}^{\frac{d}{2}+1} \,
(2\pi m_1 k_{\rm B}{T})^{\frac{d}{2}} }{n_1(\vec{r}) \;
(\Delta r\Delta p)^{d}} \; .
\label{local.entropy.density.ST}
\ee
expected from the Sackur-Tetrode formula where $\Delta r\, \Delta
p = 2\pi\hbar$ \cite{Pa72}. This result shows that our definition
of entropy corresponds to the well-known physical entropy.
Moreover, the entropy production (\ref{entrprod-p}) reduces to the form
\be
\sigma_{\rm irr} \simeq k_{\rm B}\;
\bar{\cal D} \; \frac{(\vec{\nabla} n_1)^2}{n_1}
\geq 0 \; ,
\label{entrprodaver}
\ee
with the mean diffusion coefficient
\be
\bar{\cal D} = \frac{\int_0^{\infty}\eta({\epsilon})\,
{\rm e}^{-\epsilon/k_{\rm B} {T}} {\cal
D}(\epsilon) \; d\epsilon}{\int_0^{\infty}\eta(\epsilon)\,{\rm
e}^{-\epsilon/k_{\rm B} {T}}\;
d\epsilon} \; .
\label{diffaver}
\ee
With the assumption (\ref{distrib-macro}),
Eqs. (\ref{entrprodaver}) and (\ref{diffaver})
thus give the values of entropy production and diffusion coefficient
for the diffusion of light tracer particles in a dilute binary mixture at
the temperature $T$, as can be derived from the results by Chapman and Cowling
in Ref.~\cite{CC70}.
We can therefore conclude that our whole calculation is consistent
with common knowledge on irreversible thermodynamics and
the kinetic theory of binary mixtures of light and heavy particles.

\subsection{Thermodynamic equivalence with a process producing work}

We have already shown with Eqs. (\ref{Newton}) and (\ref{thermoNewton})
that, in the random Lorentz gas,
the thermodynamic force appears as a Newtonian force acting on statistical
collections of particles. In the present subsection, our aim is to give a
general
argument demonstrating that the entropy production of diffusion
in the Lorentz gases has a physical meaning,
since it can be associated with the performance of work.
We respond in this way to the doubts expressed
by Cohen and Rondoni about a possible
``confusion between information theoretic and real entropy"
(Ref. \cite{CoRo02} p. 118), and the allusion
that entropy production in the Lorentz gas
could only refer to an abstract
information theoretic concept without physical consequence
(Ref. \cite{CoRo02} p. 118 \& 128).

\begin{figure}
\centerline{\epsfig{file=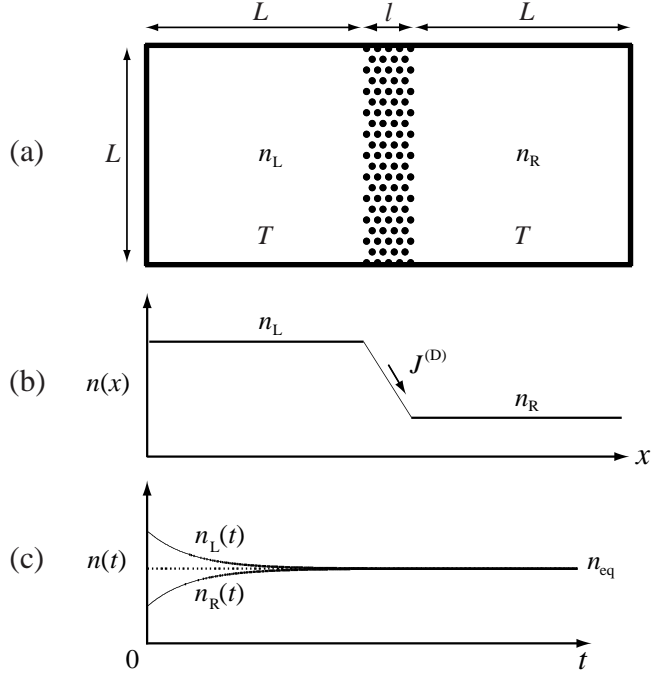,width=9cm}}
\vskip 1 cm
\caption{Irreversible process of diffusion across a Lorentz slab leading to
the equilibration of densities between both reservoirs: 
(a) configuration of the system; (b) initial density profile across the system; 
(c) time evolution of the densities which decay to their equilibrium value in each
reservoir.}
\label{fig1}
\end{figure}

Let us consider a slab of the Lorentz gas of width $l$
between two large cubic reservoirs of the
same volume $V=L^d$ containing
respectively $N_{\rm L}$ and $N_{\rm R}$ light particles
so that the particle densities
in the reservoirs are respectively
$n_{\rm L}=N_{\rm L}/V$, and $n_{\rm R}=N_{\rm R}/V$ (The subscript `L' refers
to the left-hand side and `R' to the right-hand side, see Fig.
\ref{fig1}).  We here assume that the velocity distribution of the particles
is initially taken to be a Maxwell-Boltzmann distribution at the same
temperature $T$ across the whole system, i.e.,
the distribution function has the form (\ref{distrib-macro})
{\it at the initial time} with $n_1(\vec{r})$ taking the constant
values $n_{\rm L}$ and $n_{\rm R}$ in the reservoirs and
a linear interpolating profile inbetween.
Since the diffusion process
conserves the energy of each particle, the temperature, $T$,
at the beginning is equal to the temperature after equilibrium is reached
and there is no heat produced.
However, there is a current of particles $J^{\rm (D)}$ from the high-density
reservoir to the low-density one because of the difference of
concentrations across the Lorentz slab.
The current of particles decreases
the difference of densities between both reservoirs.  As long as there is a
difference of densities, the system is
out of equilibrium, and thermodynamic equilibrium is reached when
$n_{\rm L}(t=\infty)=n_{\rm R}(t=\infty)=n_{\rm
eq}=(N_{\rm L}+N_{\rm R})/(2V)$, and the current vanishes.

The irreversible process of
equilibration by the diffusion of
particles across the Lorentz slab leads to a global production of entropy.
Indeed, in the initial situation, the
entropy of the whole system is equal to the entropies of the ideal gases
in both reservoirs if the Lorentz slab is taken to be much
smaller than the reservoirs ($l \ll L$) so that its entropy is negligible:
\be
S_{\rm initial} = S_{\rm L} + S_{\rm R} \; , \qquad \mbox{with} \quad
S_{\rm L} =  k_{\rm B}
N_{\rm L} \ln \frac{V  \gamma}{N_{\rm L}} \; , \quad \mbox{and}
\quad S_{\rm R} =  k_{\rm B} N_{\rm R} \ln \frac{V  \gamma}{N_{\rm R}} \; ,
\ee
with the constant
\be
\gamma = \frac{{\rm e}^{\frac{d}{2}+1} (2\pi m_1 k_{\rm B} T)^{\frac{d}{2}}}
{(2\pi\hbar)^d}
\ee
according to the Sackur-Tetrode formula \cite{Pa72}.
The entropy after equilibration is given by
\be
S_{\rm final} = k_{\rm B} (N_{\rm L} + N_{\rm R})  \ln \frac{2
V  \gamma}{N_{\rm L}+N_{\rm R}} \; ,
\ee
so that the total entropy increase is
\bea
\Delta S = S_{\rm final}-S_{\rm initial} &=&  k_{\rm B} N_{\rm L} \ln
\frac{2N_{\rm L}}{N_{\rm L}+N_{\rm R}} +  k_{\rm B} N_{\rm R} \ln
\frac{2N_{\rm R} }{N_{\rm L}+N_{\rm R}}  \label{entrprod1} \\
&\geq &  k_{\rm B} N_{\rm L} \left( 1 - \frac{N_{\rm L}+N_{\rm R}}{2N_{\rm
L}}\right) + k_{\rm B}
N_{\rm R} \left( 1 -
\frac{N_{\rm L}+N_{\rm R}}{2N_{\rm R}}\right) = 0 \; ,
\eea
because of the inequality $\ln(1/x)\geq 1-x$.  Accordingly, the entropy
increase (\ref{entrprod1}),
which can be rewritten in the form
\be
\Delta S =  k_{\rm B} N_{\rm L} \ln \frac{n_{\rm L}}{n_{\rm eq}} +  k_{\rm
B} N_{\rm R} \ln
\frac{n_{\rm R} }{n_{\rm eq}} \geq 0 \; ,
\label{entrprod2}
\ee
is positive in the above process without heat production.

We have now to show that the entropy
increase (\ref{entrprod2}) is
irreversibly produced by diffusion in the Lorentz slab.
Within the slab, the space and energy distributions
$f(\vec{r},\epsilon)$ of the tracer particles evolve
in time according to the diffusion equation (\ref{diff-p}).
Since the reservoirs are much larger than the Lorentz slab
($L\gg l$) the diffusion process is quasi-stationary
and a linear profile of density
$f(\vec{r},\epsilon)$ is maintained
in the slab during the whole relaxation to the equilibrium
and, this, on each energy shell.
The current (\ref{current-f}) of particles of energy $\epsilon$
is thus transverse to the slab and equal to
\be
J_f = -{\cal D} \; \frac{f_{\rm R}-f_{\rm L}}{l} \; ,
\label{current-grad}
\ee
where $f_{\rm R}$ and $f_{\rm L}$ are the densities
of particles of energy $\epsilon$ in the reservoirs and ${\cal D}={\cal
D}(\epsilon)$. These densities relax to their equilibrium values according to
the equations
\be
\frac{df_{\rm L}}{dt} = - \frac{df_{\rm R}}{dt} = -\frac{J_f}{L} \; .
\label{dfdt}
\ee
The entropy of the total system is the sum of the entropies
of both reservoirs plus the negligible entropy of the Lorentz slab.
This total entropy can be expressed in terms of the entropy density
$s$ which obeys the balance equation
(\ref{bal.entropy})
so that the time derivative of the total entropy is given by
\be
\frac{dS}{dt} = A \int dx \; \partial_t \, s
= A \int dx \; (-\partial_xJ_s+\sigma_{\rm irr}) \; ,
\ee
where $A=L^{d-1}$ is the area of a section of both reservoirs
perpendicular to the direction of the current (see Fig. \ref{fig1}).
The entropy production vanishes in both reservoirs
and takes the positive value (\ref{entrprod-p}) in the Lorentz slab.
Since the total system is isolated, the entropy current vanishes
at its borders so that the only contribution to the entropy variation
comes from the irreversible entropy production (\ref{entrprod-p})
in the Lorentz slab:
\be
\frac{dS}{dt} = A \int dx \; \sigma_{\rm irr}
= A \int_L^{L+l} dx \int_0^{\infty} d\epsilon \; k_{\rm B} \; {\cal D} \;
\frac{(\partial_xf)^2}{f} \geq 0
\; .
\label{pre-dSdt}
\ee
In a quasi-stationary evolution, the density of particles of energy $\epsilon$
is linear in the slab
\be
f=f_{\rm L} + \frac{f_{\rm R}-f_{\rm L}}{l} \, (x-L) \; ,
\qquad \mbox{for} \quad L<x<L+l \; ,
\ee
so that the entropy variation (\ref{pre-dSdt}) becomes
\be
\frac{dS}{dt} = k_{\rm B} \, A \int_0^{\infty} d\epsilon \; {\cal D}
\; \frac{f_{\rm R}-f_{\rm L}}{l} \; \ln \frac{f_{\rm R}}{f_{\rm L}} \geq 0 \; .
\ee
According to Eq. (\ref{current-grad}), the gradient of densities
can be expressed in terms of the current whereupon
\be
\frac{dS}{dt} = - k_{\rm B} \, A \int_0^{\infty} d\epsilon
\; J_f \, \ln \frac{f_{\rm R}}{f_{\rm L}} \; .
\ee
Besides, Eq. (\ref{dfdt}) allows us to write the current
as the time derivatives of the densities and we get
\be
\frac{dS}{dt} = - k_{\rm B}  V \int_0^{\infty} d\epsilon
\left( \frac{df_{\rm L}}{dt} \; \ln f_{\rm L}
+ \frac{df_{\rm R}}{dt} \; \ln f_{\rm R} \right) \; ,
\label{dSdt}
\ee
where $V=AL$ is the volume of each reservoir.
Eq. (\ref{dSdt}) can now be integrated over time to obtain
\be
\Delta S = S_{\rm final} - S_{\rm initial} =  k_{\rm B} V \int_0^{\infty}
d\epsilon
\left( f_{\rm L} \; \ln \frac{2\, f_{\rm L}}{f_{\rm L}+f_{\rm R}}
+f_{\rm R} \; \ln \frac{2\, f_{\rm R}}{f_{\rm L}+f_{\rm R}}
\right) \geq 0 \; ,
\label{DS-f}
\ee
where $f_{\rm L}$ and $f_{\rm R}$ denote the {\it initial}
distributions given by the Boltzmann distribution (\ref{distrib-macro})
with the {\it initial} densities $n_{\rm L}=N_{\rm L}/V$ and
$n_{\rm R}=N_{\rm R}/V$, while the final distributions are equal to
$(f_{\rm L}+f_{\rm R})/2$ in both reservoirs.  The integral over the energy
can be
performed in Eq. (\ref{DS-f}) in the same way as the Sackur-Tetrode entropy
(\ref{local.entropy.density.ST}) was obtained. Consequently, we recover the
entropy
increase (\ref{entrprod2}) demonstrating that this entropy is irreversibly
produced by diffusion in the Lorentz slab.

\begin{figure}
\centerline{\epsfig{file=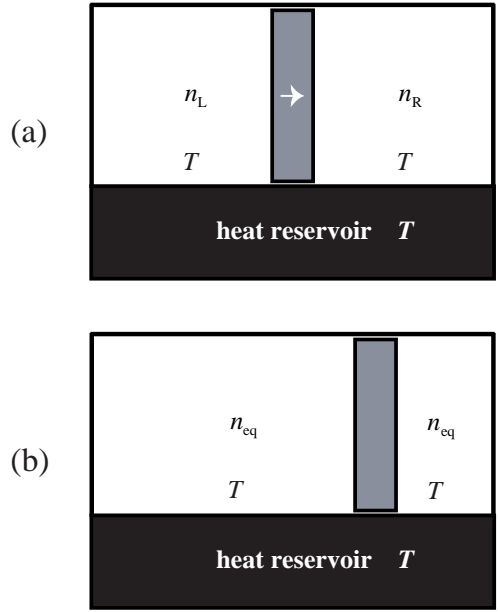,width=7cm}}
\vskip 1 cm
\caption{Replacement of the Lorentz slab by a movable piston so that work
can be extracted from the difference
of densities if the whole system is in contact with a heat reservoir: (a)
initial
configuration of the system; (b) final configuration after the work $W$ has
been extracted.}
\label{fig2}
\end{figure}

Now, the entropy increase (\ref{entrprod2}) corresponds to some work
which could otherwise be used if we were to
replace the Lorentz slab by a movable piston of equal volume, and if the
system were put in contact with a
heat reservoir (see Fig. \ref{fig2}).  In the initial situation where there
is a difference of particle
densities between both sides, the piston is submitted to a force due to the
difference of pressures
\be
P_{{\rm L},{\rm initial}} = \frac{N_{\rm L}k_{\rm B} T}{V} \; , \qquad
\mbox{and}
\qquad P_{{\rm R},{\rm initial}} = \frac{N_{\rm R}k_{\rm
B} T}{V} \; .
\ee
This force can produce work if the piston is slowly moved until both
pressures
equilibrate.  We suppose that the piston is moved so slowly that the
process is isothermal.  The equilibrium of
pressures is reached when the high-density, high-pressure fluid has
expanded and compressed the fluid on
the other side so that both have the same equilibrium density $n_{\rm eq}$,
the temperature $T$ being kept
constant.  The volumes of both fluids have changed so that
\be
n_{\rm eq} = \frac{N_{\rm L}}{V_{\rm L}} = \frac{N_{\rm R}}{V_{\rm R}}=
\frac{N_{\rm L}+N_{\rm R}}{2V} \; , \qquad
\mbox{and} \qquad
P_{{\rm L},{\rm final}} = \frac{N_{\rm L}k_{\rm B} T}{V_{\rm L}} = P_{{\rm
R},{\rm
final}} = \frac{N_{\rm R}k_{\rm B} T}{V_{\rm R}} \; ,
\ee
with $V_{\rm L}+V_{\rm R}=2V$, the numbers of particles being here constant
in each fluid.
The work extracted in this process is
\be
W = \int_{\rm initial}^{\rm final} \left(P_{\rm L} \; dV_{\rm L} + P_{\rm
R} \; dV_{\rm R}\right)
= k_{\rm B} T \left( N_{\rm L} \ln \frac{V_{\rm L}}{V} + N_{\rm R} \ln
\frac{V_{\rm R}}{V} \right) \; .
\ee
Since the ratios of volumes are equal to the ratios of densities
\be
\frac{V_{\rm L}}{V}= \frac{n_{\rm L}}{n_{\rm eq}} \; ,
\qquad \mbox{and}\qquad
\frac{V_{\rm R}}{V}= \frac{n_{\rm R}}{n_{\rm eq}} \; ,
\ee
we find that the extracted work is equal to the temperature multiplied by
the entropy increase during the
previous process:
\be
W = k_{\rm B} T \left( N_{\rm L} \ln \frac{n_{\rm L}}{n_{\rm eq}} +  N_{\rm
R} \ln
\frac{n_{\rm R} }{n_{\rm eq}}\right) = T \; \Delta S \; .
\ee
Since the initial and final energies of the particles in both fluids are
equal by the equation of state $E=(d/2)Nk_{\rm B}T$ for ideal gases,
the work $W$ has been pumped from the heat reservoir.
Accordingly, we conclude that the entropy
(\ref{entrprod2}) irreversibly produced by diffusion in the Lorentz slab
corresponds to some work which is lost during the process of diffusion.  In
this sense, merely by randomizing the velocity directions of the particles, 
the diffusion process dissipates an energy which
could otherwise be used, which confirms that
diffusion in the Lorentz gas is an irreversible process obeying the Second
Law of thermodynamics.


\section{The irreversible thermodynamics of the multibaker model of diffusion}
\label{multibaker}

In this section, we show that the previous considerations for the diffusive
Lorentz gas extend to the multibaker
model of diffusion as well, contrary to
the statement by Cohen and Rondoni
that ``there
is no multibaker-analog of the concept of local thermodynamic equilibrium"
(Ref.
\cite{CoRo02}, p. 125).

\subsection{The multibaker map as a model of diffusion}

The multibaker model can be considered to be a simplification of the Birkhoff
map of the
hard-disk Lorentz gas.  The Birkhoff map of the Lorentz gas is the map
induced by the collisional dynamics
between the disk scatterers at constant energy.  This map is constructed in
terms of two Birkhoff coordinates:
(i) the angle $0\leq\theta<2\pi$ of the position of collision on the disk,
and (ii) the canonically conjugate variable
which is the component $-1<\varpi=\sin\psi<+1$ of velocity that is
tangent to the disk, $\psi$ being the
angle between the velocity and the normal to the disk at collision.  The
kinetic energy $\epsilon$ is conserved
during the whole motion so that the trajectories at different energies are
similar after a rescaling of the time
in terms of the speed $v(\epsilon)=\sqrt{2\epsilon/m_1}$.  Between two
collisions, the Birkhoff coordinates
$\xi=(\theta,\varpi)$ are constant while the path length $\lambda$
increases linearly with time.  Furthermore,
the lattice vector $\vec{\ell}$ labeling the disk on which
the collision has occurred is also constant between two collisions.
Accordingly, between two collisions,
the equations of motion become:
\begin{eqnarray}
\frac{d\xi}{dt} &=& 0 \; , \\
\frac{d\lambda}{dt} &=& v(\epsilon) = \sqrt{\frac{2\epsilon}{m_1}} \; ,\\
\frac{d\epsilon}{dt} &=& 0 \; , \\
\frac{d\vec{\ell}}{dt} &=& 0 \; .
\end{eqnarray}
At collisions, the Birkhoff coordinates, the path length, as well as the
lattice vector, $\vec{\ell}$, jump
according to the map
\begin{eqnarray}
\xi_{n+1} &=& \phi(\xi_n) \; ,\\
\lambda_{n+1} &=& \lambda_n + \Lambda(\xi_n) \; ,\\
\epsilon_{n+1} &=& \epsilon_n \; ,\\
\vec{\ell}_{n+1} &=& \vec{\ell}_n+\vec{L}(\xi_n) \; ,
\end{eqnarray}
where $\Lambda(\xi)$ is the path length of a free flight issued from the
impact point $\xi$ to the next
collision.  The lattice vector $\vec{L}(\xi)$ gives the disk on which the
next collision occurs.  The time of
flight between two collisions is given by
$T(\xi,\epsilon)=\Lambda(\xi)/v(\epsilon)$, which defines the
first-return time function or ceiling function.  The Birkhoff map together
with the first-return time function
provide an equivalent description of the trajectories of the
continuous-time dynamics in the Lorentz gas at a
given energy.  Each phase-space point on a segment of trajectory between
two collisions is uniquely represented
by the coordinates $(\vec{\ell},\theta,\varpi,\lambda,\epsilon)$.  The
hard-disk Lorentz gas has two degrees of
freedom so that its phase space is four-dimensional.  The four original
variables of position and momentum $(\vec{r},\vec{p})$ are
here replaced by the four continuous variables
$(\theta,\varpi,\lambda,\epsilon)$, the lattice vector $\vec{\ell}$
taking its values in a discrete set.

\begin{figure}
\centerline{\epsfig{file=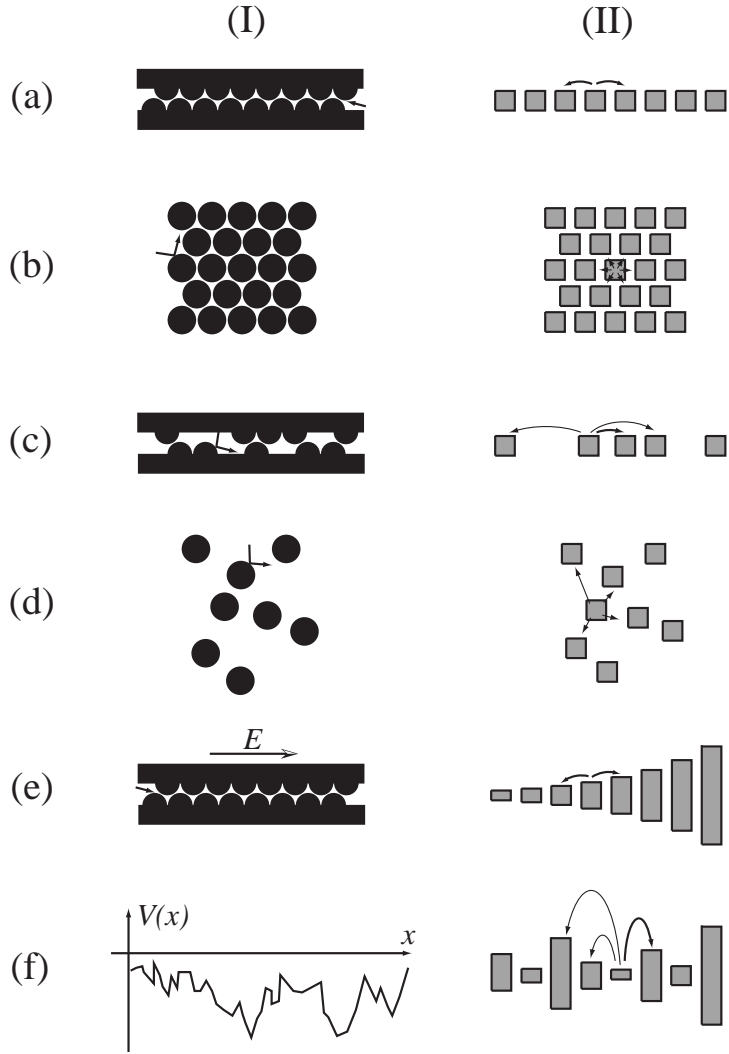,width=10cm}}
\vskip 0.2 cm
\caption{Schematic representation of different Lorentz gases (I) and their
corresponding multibaker model (II)
for configurations of scatterers which are: (a) one-dimensional and
periodic; (b) two-dimensional and periodic;
(c)  one-dimensional and disordered; (d) two-dimensional and disordered;
(e) with a constant electric field $E$;
(f) with a disordered potential $V(x)$. In each example of column (II), the
multibaker map
acts on the grey squares or rectangles by a baker-type transformation
combined with
translations (depicted by arrows) of the cut pieces to the neighboring
squares or
rectangles. The multibaker (IIa) is defined by Eqs. (\ref{dyadic}) and
(\ref{jump}).}
\label{fig3}
\end{figure}

In the multibaker map, the Birkhoff map is simplified by considering a
baker-type transformation instead of
the nonlinear transformation $\phi$ induced by the collisional dynamics
from disk to disk.  Different multibaker
models have been constructed corresponding to different Lorentz models (see
Fig. \ref{fig3})
\cite{Ga92,Ga98,TeVoBr96,TaGa95,TaGa99,TaGa00,Ra99,TeVo00,VoTeMa00,Vo02}.

A dyadic version of the multibaker map is defined by the following mapping
which acts on a chain of unit squares
\cite{TaGa95}
\be
\phi(x,y) \ = \ \cases{  \Bigl( 2x,{y \over 2}\Bigr) \ , \qquad \qquad \ \
0 \leq x
\leq {1\over 2} \; ,  \cr
\Bigl(2x-1,{{y+1}\over 2}\Bigr) \ , \qquad {1\over 2} < x
\leq 1 \; ,  \cr}
\label{dyadic}
\ee
with the jump vector
\be
L(x,y) \ = \ \cases{  -\alpha \; , \qquad 0 \leq x
\leq {1\over 2} \; ,  \cr
+\alpha \; , \qquad {1\over 2} < x \leq 1 \; ,  \cr}
\label{jump}
\ee
where $\alpha$ is the spatial distance separating two neighboring squares.
The dynamics of the multibaker is time-reversal symmetric under the involution
$I(x,y)=(1-y,1-x)$ so that $\phi^{-1}=I\circ\phi\circ I$.  The dyadic
multibaker (\ref{dyadic}) is
hyperbolic with the positive Lyapunov exponent $\ln 2$ and chaotic with the
KS entropy $h_{\rm
KS}=\ln 2$.  It is also area-preserving as is the Birkhoff
map, and the coordinates $(0<x<1,0<y<1)$ play a similar role as the
Birkhoff coordinates $(\theta,\varpi)$.  They also form a pair of
canonically conjugate variables.  On the other hand, the
coordinate $x$ corresponds to the unstable direction of the mapping and the
coordinate $y$ to its stable
direction.  This is not the case for the Birkhoff coordinates
$(\theta,\varpi)$ which mix the stable and unstable
directions of the Birkhoff mapping. Therefore, if the linearization of the
Birkhoff map of the hard-disk Lorentz
gas were possible, a change of variables would be required to relate the
Birkhoff coordinates $(\theta,\varpi)$
to the multibaker coordinates $(x,y)$.  The Birkhoff map of the hard-disk
Lorentz gas has
singularity lines connected to grazing collisions
which are absent in the multibaker map.
  In this regard,
the multibaker map constitutes an
important simplification of the Birkhoff map of the hard-disk dynamics but
the analogy is complete as far as all
the other properties are concerned, such as the dimensionality of the phase
space, the diffusivity of the
particles moving under the multibaker dynamics, the area-preserving
property, as well as the positivity of the
Lyapunov exponent.  It is possible to take into account the energy of the
particle in the multibaker model by
simply keeping this variable as in the Lorentz gas as explained in Refs.
\cite{TaGa99,TaGa00,Vo02}.  In analogy with
the Lorentz gas, energy is conserved.  Moreover, we may suppose that the
path length between two successive
intersections in the plane $(x,y)$ of the multibaker map is independent of
the coordinates $(x,y)$ and takes
the constant value $\Lambda(x,y)=\Lambda_c$.

In the multibaker map, the phase space is based on a chain of squares.
Each square of it is stretched, cut in
two pieces, and mapped on the neighboring squares.
The multibaker is mixing,
as is the hard-disk Lorentz gas so that a {\it local equilibrium} is
established in
each square of the chain because the dynamics randomizes the coordinates
$(x,y)$.
Moreover, this mapping
induces jumps of the particle
along the chain of squares.  The jumps generate a symmetric random walk
with diffusion coefficient
${\cal D}=v(\epsilon) \alpha/2$, where $v(\epsilon)$ is the speed of the
particles which may be supposed to be
the same as in the hard-disk Lorentz gas.

\subsection{The thermodynamics of the multibaker map}

Except that the dynamics is simplified, the physical interpretation of the
multibaker map of diffusion is
similar to the one of the corresponding Lorentz gas. Thus the
thermodynamic considerations of the previous
section apply {\it mutatis mutandis}.  Let us consider a gas of $N_1$
particles in a multibaker chain containing
$N_2$ squares.  We can impose periodic boundary conditions at
the ends of the chain for instance.
The phase space of the $N_1$ particles is
\be
{\cal M}=\{
(\ell_1,x_1,y_1,\lambda_1,\epsilon_1,...,
\ell_{N_1},x_{N_1},y_{N_1},\lambda_{N_1},\epsilon_{N_1})\vert
\ell_1+x_1<...<\ell_{N_1}+x_{N_1}\} \; ,
\ee
with $\ell_j \in \{1,2,3,...,N_2\}$,
$0<x_j<1,0<y_j<1,0<\lambda_j<\Lambda_c,0<\epsilon_j$,
with, again, the ordering required by the indistinguishability of the
particles.
In this phase space, the equilibrium microcanonical
probability distribution has the
following density function
\be
F(x_1,y_1,\lambda_1,\epsilon_1,...,
x_{N_1},y_{N_1},\lambda_{N_1},\epsilon_{N_1})
= \cases{
\frac{\displaystyle N_1!}{\displaystyle
{\cal V}^{N_1} \Delta\epsilon^{N_1}} \qquad{\rm for}
\quad \epsilon<\epsilon_j<\epsilon+\Delta\epsilon \; , \cr\cr
0 \qquad {\rm otherwise} \; , \cr}
\ee
where $\cal V$ is the volume of the domain where the light particles move.
This volume can
be supposed to depend on the physical volume $V$ and on the number $N_2$ of
hard disks as in
the corresponding Lorentz gas.  The equilibrium Gibbs entropy
(\ref{Gibbs.entropy})
calculated for this distribution function is given by
\be
S = k_{\rm B} N_1 \; \ln \frac{ {\cal V} \; {\rm e}}{N_1 \Delta
x\Delta y \Delta \lambda} \; ,
\ee
where $\Delta x\Delta y \Delta \lambda\Delta \epsilon$ is the volume of the
cells partitioning the phase space.
The Gibbs relation (\ref{Gibbs.relation}) is
satisfied with a vanishing inverse temperature,
the chemical potential of the light particles given by
\be
\frac{\mu_1}{\cal T}=k_{\rm B} \ln \frac{ N_1 \Delta x\Delta y \Delta \lambda
}{{\cal V} } \; ,
\ee
and another one for the scatterers
\be
\frac{\mu_2}{\cal T} = \frac{N_1 k_{\rm B}}{\cal V}
\left(-\frac{\partial{\cal V}}{\partial
N_2}\right) \; .
\ee
Hence, the considerations of the previous section follow, showing that
the multibaker map is also consistent with the irreversible
thermodynamics of diffusion processes, contrary to the claims by Cohen and
Rondoni \cite{CoRo02}.

In Refs.
\cite{TaGa99,TaGa00,GiDo99,TeVo00,VoTeMa00,TeVoMa01,MaTeVo01,Vo02,Ga97b,GiDoGa00},
the entropy production
associated with the transport processes
sustained by multibaker models was derived and the expressions
expected from
irreversible thermodynamics have been obtained. More recently,
the {\it ab initio}
derivation of entropy production of diffusion has been extended
to the continuous-time dynamics of the
Lorentz gas as well as to many-body
systems composed of a tracer particle moving in a fluid of other particles
\cite{DoGaGi02}.  These works show that the
thermodynamically expected entropy production can be successfully
derived from dynamics in
the Lorentz gases and multibaker maps.


\section{Conclusions}
\label{conclusions}

This paper provides counter-arguments refuting the doubts and
criticisms expressed by Cohen and Rondoni concerning the thermodynamic
considerations of Lorentz gases and multibaker maps.  We have shown that
their claims cannot be taken uncritically, and do not stand up under
a rather direct consideration of the irreversible thermodynamics of
isothermal mixing of distinct chemical species. We hope that this
exchange of ideas will be helpful for clarifying 
the thermodynamics of the model systems we study. 
We feel that, at best, the arguments of Cohen and Rondoni should be
considered as restatements of the facts that a
study of simple systems may not reveal all of the features and difficulties that
one encounters in a study of complicated ones.
Yet, whatever the specificities of simple systems may be,
we have seen that they remain within the realm of irreversible thermodynamics.
The present work thus shows that irreversible thermodynamics indeed has a
universal applicability in the class of 
spatially-extended many-particle systems.

We should point out that the goal of the line of research presented in the
papers criticized by Cohen and Rondoni is not the treatment of entropy
production in Lorentz gases or multi-baker maps. Rather, these papers represent
steps on the way to a treatment of entropy production in classical
fluids. As is often the case in statistical physics, the study of
simple models allows workers to develop insights into the
ideas and techniques needed to treat more realistic systems. From this
point of view the criticisms of Cohen and Rondoni, while mistaken, as
we argue here, would also be irrelevant if
our line of research shows that similar methods can be useful in the
study of fluid systems where all of the particles can interact with
each other. In fact, we have shown in a previous paper \cite{DoGaGi02},
that the methods under discussion apply equally well to a periodic,
fluid system with a tracer particle. In this case, we considered a
fully interacting, periodic $N+1$ particle system, in a microcanonical
ensemble. For this system the presence of interacting, 
moving particles does not change the arguments in any important way.

Cohen and Rondoni have also raised more technical objections to
our use of Sinai-Ruelle-Bowen measures to calculate the rate of
entropy production as described in Refs.
\cite{Ga98,GiDo99,Ga97b,GiDoGa00,DoGaGi02}
which also require a response. This will be supplied in a
future publication.

\vskip 0.5 cm

{\bf Acknowledgments.}
The authors thank Doron Cohen, E. G. D. Cohen, T. Gilbert, H.~Larralde,
L. Rondoni, H. van Beijeren, J. Vollmer, and D. Wojcik for
interesting and helpful discussions.
PG thanks the National Fund for Scientific Research (FNRS Belgium) for
financial support. JRD thanks the National Science Foundation (USA)
for support under grant PHY-98-20824.


\end{document}